\documentclass[apj]{emulateapj}

\usepackage{amsmath}	% Advanced maths commands
\usepackage{amssymb}	% Extra maths symbols
\usepackage{multirow}
\usepackage{subfigure}
\usepackage{booktabs}
\usepackage{color}
\def\wise{\textit{WISE} }
\def\madcows{MaDCoWS }
\shorttitle{Radio Morphologies of \madcows AGN}
\shortauthors{E. Moravec et al.}

\begin{document}

\title{The Massive and Distant Clusters of WISE Survey V: Extended Radio Sources in Massive Galaxy Clusters at $z \sim 1$}

\author{Emily Moravec\altaffilmark{1},
Anthony H. Gonzalez\altaffilmark{1}, 
Daniel Stern\altaffilmark{2},
Mark Brodwin\altaffilmark{3},
Tracy Clarke\altaffilmark{4},
Bandon Decker\altaffilmark{3},
Peter R. M. Eisenhardt\altaffilmark{2},
Wenli Mo\altaffilmark{1}, 
Christine O'Donnell\altaffilmark{5}, 
Alexandra Pope\altaffilmark{6},
Spencer A. Stanford\altaffilmark{7}, 
and
Dominika Wylezalek\altaffilmark{8}
}

\affil{$^1$Department of Astronomy, University of Florida, Gainesville, FL, 32611, USA}
\affil{$^2$Jet Propulsion Laboratory, California Institute of Technology, Pasadena, CA 91109, USA}
\affil{$^3$Department of Physics and Astronomy, University of Missouri, 5110 Rockhill Road, Kansas City, MO 64110, USA}
\affil{$^4$Naval  Research  Laboratory,  Code  7213,  Washington,  DC,
20375, USA}
\affil{$^5$Department of Astronomy, University of Arizona, 933 North Cherry Avenue, Tucson, AZ 85721 }
\affil{$^6$Department of Astronomy, University of Massachusetts, 710 North Pleasant Street Amherst, MA 01003-9305}
\affil{$^7$Department of Physics, University of California, Davis, One Shields Avenue, Davis, CA 95616, USA}
\affil{$^8$European Southern Observatory, Karl-Schwarzschild-Str. 2, 85748 Garching bei M\"unchen, Germany}

%%%%%%%%%%%%%%%%%%%%%%%%%%%%%%%%%%%%%
% Abstract of the paper
\begin{abstract}
We present the results from a pilot study with the Karl G. Jansky Very Large Array (JVLA) to determine the radio morphologies of extended radio sources and the properties of their host-galaxies in 10 massive galaxy clusters at $z \sim 1$, an epoch in which clusters are assembling rapidly. These clusters are drawn from a parent sample of \textit{WISE}-selected galaxy clusters that were cross-correlated with the VLA Faint Images of the Radio Sky at Twenty-Centimeters survey (FIRST) to identify extended radio sources within 1$^{\prime}$ of the cluster centers. Out of the ten targeted sources, six are FR II sources, one is an FR I source, and three sources have undetermined morphologies. Eight radio sources have associated \textit{Spitzer} data, 75\% presenting infrared counterparts. A majority of these counterparts are consistent with being massive galaxies. The angular extent of the FR sources exhibits a strong correlation with the cluster-centric radius, which warrants further investigation with a larger sample.
\end{abstract}

\maketitle

%%%%%%%%%%%%%%%%%%%%%%%%%%%%%%%%%%%%%%%%%%%%%%%%%%
%%%%%%%%%%%%%%%%% BODY OF PAPER %%%%%%%%%%%%%%%%%%
\section{Introduction}\label{intro}
Two major modes of Active Galactic Nuclei (AGN) feedback have been identified and differentiated by the energy outflow near the black hole: `quasar-mode' and `radio-mode' (see \citealt{Fabian12} for a review).~During quasar-mode feedback (also known as radiative or wind mode) the AGN is radiatively efficient and its outflows can expel interstellar gas from the host galaxy, slowing the infall of matter into both the galaxy and the central Supermassive Black Holes \citep[SMBH,][]{Kauffmann00, Granato04, DiMatteo05, Hopkins10}. On the other hand, during radio-mode feedback (also known as kinetic, radio-jet, or maintenance mode) the AGN is radiatively inefficient and is capable of driving powerful, kpc-scale jets. Despite this prevailing nomenclature, there also exists a substantial population of radio-loud quasars, accounting for 10-15\% of the luminous quasar population, where the central engine is both radiatively efficient and driving a powerful, kpc-scale jet \citep[e.g.,][]{Stern00}. 

Radio AGN and dense environments seem to be connected in a myriad of ways. During radio-mode feedback, the radio jets of AGN can have a significant influence on their surrounding medium and environment, such as in galaxy clusters. The epoch $1<z<2$ is an important epoch for large-scale structure formation during which clusters are assembling rapidly and the intracluster medium (ICM) is forming. Recent observations indicate that AGN play a crucial role in regulating the cooling of the ICM even up to $z \sim$ 1 \citep[]{McDonald13, HL15}. The hot, X-ray emitting ICM in the central region of massive galaxy clusters often has a radiative cooling time shorter than the Hubble time. In the absence of any heating, the ICM should therefore have had time to cool, condense, and produce large flows of cooling material. However, X-ray observations with \textit{Chandra} and \textit{XMM-Newton} have shown that there is considerably less cooling material than predicted by standard cooling-flow models \citep[e.g.,][]{Bohringer01, Tamura01, Peterson01, Peterson03, Peterson06, McN07}. This discrepancy between the cooling-flow models and the observations is traditionally known as the ``cooling flow problem." 

The leading mechanism proposed to rectify this discrepancy is feedback from the central AGN associated with the brightest cluster galaxy (BCG). The AGN can inflate low density X-ray cavities via radio-emitting jets. These jets are thought to limit central star formation and prevent the hot X-ray gas from cooling by transferring mechanical energy to the ICM through shock and sound waves \citep[]{Fabian00, Birzan04, Dunn06, Croton06, Fabian06, Sanders08, Birzan08,  Dong10, Cavagnolo10, OSul11} producing the observed ``entropy floor" of $\sim$ 10 keV cm$^2$ in clusters \citep[e.g.,][]{Fabian12, McN12}.

The connection between radio-emitting AGN and their environments is showcased in other ways beyond the observed interaction between radio jets and the ICM in cooling core clusters. The association of radio galaxies with galaxy clusters dates back to the 1950's, and \cite{Minkowski60} established a redshift record of z=0.46 for 3C 295 that stood for 15 years. More recently, there have been successful searches for rich, high redshift ($z\gtrsim1$) clusters, cluster candidates, and protoclusters using radio-AGN such as the Clusters Around Radio-Loud AGN program \citep[CARLA,][]{Wylezalek13,Wylezalek14,Noirot16,Noirot18}, the Clusters Occupied by Bent Radio AGN \citep[COBRA,][]{PM17}, and \cite{Castignani14} and \cite{Rigby14} which use high-redshift radio galaxies.

One way to further probe the role of radio-mode AGN in the cluster environment is investigating the morphology of extended radio emission from AGN in galaxy clusters. Double-lobed radio sources are a prevalent radio morphology and are divided into two morphological classes: Fanaroff and Riley (FR) classes I and II \citep[]{FR74}. FR I sources are `edge-darkened' in appearance in that the emission is brighter near the radio core and becomes fainter radially outward. FR II sources are `edge-brightened' in appearance in that the well-separated lobes end in distinctive areas of brightest emission (i.e. ``hotspots"). Besides the morphological distinctions, there is a separation in radio power between the two classes that occurs at P$_{1.4}\sim 10^{25}$ W Hz$^{-1}$, where P$_{1.4}$ is the radio luminosity at rest-frame 1.4 GHz. FR II sources are generally above this threshold and FR I sources are generally below this threshold \citep[]{FR74,LO96}, though this division is not absolute.

The radio morphology of a source has been shown to be an effective probe of its environment. For example, studies showed that FR I sources reside in richer environments than FR II sources \citep[]{Longair79, PP88, LO96, Miller99, Wing11, Gendre13}. Additionally, the environments of FR II sources are richer at higher redshift than at lower redshift \citep[]{Best00}, whereas the environments of FR I sources do not seem to change with cosmic epoch as they remain dense and constant with time \citep[]{Hill91, Zirbel97, Stocke99, Fujita16}. Lastly, bent-tail radio sources have been used to calculate the density of the surrounding medium \citep{Freeland08,Freeland11}. These relationships between radio galaxy morphology and their environment demonstrate how the morphology of a radio source can be used as a probe to understand the evolution of and interaction between radio galaxies and their environment.

In this paper, we conduct a pilot study of a sample of extended radio sources associated with clusters from the Massive and Distant Clusters of \wise Survey \citep[MaDCoWS,][Mo et al. in press]{Stanford14, Brodwin15, Gonzalez15, G18}. We morphologically classify extended radio sources in 10 clusters, use these morphologies to place constraints on the physical characteristics of the jets, and investigate the infrared and optical counterparts to the radio sources with the goal to better understand the role of AGN in the cluster environment and the cluster environment as a whole. Throughout this paper, we adopt the flat $\Lambda$CDM cosmological model with a \cite{Planck15} cosmology, $H_0$ = 67.8 km s$^{-1}$, $\Omega_{m}$ = 0.308, $\Omega_{\Lambda}$ = 0.692, and $n_s$ = 0.968. Unless otherwise noted, all magnitudes are on the Vega system. 

\section{Sample Selection and Observations} \label{sample}

\begin{deluxetable*}{ccllccccc}
	\tablecaption{Cluster Properties and JVLA Observations\label{tb:vla_obs}}
	\tablehead{\colhead{Cluster} & \colhead{$z_{\rm{phot}}$} & \colhead{RA} & \colhead{Dec} & \colhead{Obs. Date} & \colhead{Int. Time} & \colhead{Phase Cal.} & \colhead{Flux Cal.} & \colhead{RMS} \\ & & & & & \colhead{(s)} & & & \colhead{($\mu$Jy/beam)}}
	\startdata
	MOO J0015+0801 & 0.9$\pm{0.04}$ & 00:15:23.98 & +08:01:31.8 & 2016 October 26 & 1114 & J0022+0608 & 3C 48 & 45 \\
	MOO J0121$-$0145 & 0.98$\pm{0.07}$ & 01:21:51.94 & $-$01:45:44.0 & 2017 January 18 & 1152 & J0125$-$0005 & 3C 48 & 24 \\
	MOO J0228$-$0644 & 0.86$^{+0.09}_{-0.06}$ & 02:28:00.84 & $-$06:44:52.6 & 2107 January 07 & 1152 & J0241$-$0815 & 3C 48 & 30\\
	MOO J0250$-$0443 & 1.06 & 02:50:43.04 & $-$04:43:21.8 & 2017 January 12 & 1152 & J0241$-$0815 & 3C 48 & 20\\
	MOO J0300+0124 & 1.33$^{+0.04}_{-0.06}$ & 03:00:10.71 & +01:24:49.8 & 2016 December 05 & 1112 & J0323+0534 & 3C 138 & 33\\
	MOO J1358+2158 & 0.99$^{+0.07}_{-0.06}$ & 13:58:25.47 & +21:58:47.4 & 2017 January 23 & 1152 & J1357+1919 & 3C 286 & 31\\
	MOO J1412+4846 & 1.06 & 14:12:57.69 & +48:46:01.0 & 2017 January 23 & 1152 & J1423+4802 & 3C 286 & 30\\
	MOO J1435+4759 & 1.02$^{+0.06}_{-0.05}$ & 14:35:23.47 & +47:59:41.1 & 2017 January 23 & 1152 & J1423+4802 & 3C 286  & 27\\
	MOO J1731+5857 & 1.02$\pm{0.08}$ & 17:31:04.91 & +58:57:50.1 & 2016 December 30 & 1152 & J1756+5748 & 3C 386 & 19\\
	MOO J2247+0507 & 1.02$\pm{0.05}$ & 22:47:15.62 & +05:07:49.1 & 2016 November 23 & 1152 & J2257+0743 & 3C 48 & 49
	\enddata
	\tablecomments{The prefix of the cluster name stands for Massive Overdense Object. We note that MOO J0250$-$0443 and MOO J1412+4749 lack IRAC data for the calculation of a photometric redshift, thus we assume the median redshift of the MaDCoWS survey for these clusters \citep[$z=1.06$,][]{G18}. }
\end{deluxetable*}

\subsection{Sample}
We construct our sample by cross-correlating two surveys: \madcows and the Faint Images of the Radio Sky at Twenty-cm \citep[FIRST,][]{Becker94,Becker95} Survey. The primary \madcows search covers 17,668 deg$^2$ of the extragalactic sky at $\delta$ $>$ $-$30$^{\circ}$ using a combination of data from the Wide-field Infrared Survey Explorer \citep[\textit{WISE}:][]{Wright10} and the Panoramic Survey Telescope and Rapid Response System \citep[Pan-STARRS:][]{PS16} to detect cluster candidates in the redshift range of 0.8 $\lesssim z \lesssim$ 1.4. The cluster center coordinates listed in Table \ref{tb:vla_obs} and used in this analysis (referred to as the cluster center) correspond to catalog coordinates from the original \textit{WISE}---PanSTARRS search as described in \cite{G18}.

Mo et al. (in press) find that 19\% of the \madcows clusters within the FIRST footprint have at least one FIRST source coincident with the inner $1\arcmin$ region. From the 1300 highest significance \madcows clusters in the FIRST footprint, we identified a parent sample of 51 clusters with extended radio sources. These clusters satisfy the criteria of having FIRST sources with deconvolved sizes exceeding 6.5$^{\prime\prime}$ (50 kpc at $z \simeq 1$) within 1$^{\prime}$ of the cluster center. These were the primary sources targeted for JVLA follow-up observations, the first ten of which were observed in the 2016B observing semester and are presented in this work (see Table \ref{tb:vla_obs} and \S\ref{sect:vla}). In some cases, there were additional FIRST sources within $\sim$ 1$^{\prime}$ of the cluster center. We refer to these sources as the secondary sources and discuss them in the Appendix.

Since there are no spectroscopic redshifts available to confirm cluster membership, there is a possibility that the extended sources within 1$^{\prime}$ of the cluster center are chance superpositions. To determine the probability that these sources are interlopers, we first calculate the field density of extended sources with sizes $\gtrsim$ 6.5$^{\prime\prime}$ ($\Sigma_{\rm f}$). We then compute the surface density of extended sources within 5$^{\prime}$-20$^{\prime}$ of any of the $\sim$1300 MaDCoWS clusters that lie in the FIRST footprint \citep[]{G18}. We find $\Sigma_{\rm f}$ = 2.65 $\pm$ 0.05$\times$10$^{-3}$ arcmin$^{-2}$, indicating that we should expect a total of 0.424 $\pm$ 0.008 sources ($\sim 1 \%$) to be interlopers.

\subsection{FIRST Data}
The FIRST survey covers 10,575 deg$^2$ of the North and South Galactic Caps. These data were taken using the Karl G. Jansky Very Large Array (JVLA) in the B-configuration in L-band. FIRST has a typical RMS of 0.15 mJy and a resolution of 5$^{\prime\prime}$. Table \ref{tb:morph} lists relevant properties obtained from the FIRST catalog.

We calculate the radio luminosity of the FIRST sources,
\begin{equation}\label{eqn:power}
    P_{1.4} = 4\pi{D_L}^2S_{1.4}(1+z)^{\alpha-1},
\end{equation}
where $D_L$ is the luminosity distance at the redshift of the cluster, $S_{1.4}$ is the integrated radio flux at 1.4 GHz from FIRST, (1 + $z$)$^{\alpha-1}$ includes both the distance dimming and K-correction, and $\alpha$ is the radio spectral index ($S_{\nu} \propto \nu^{-\alpha}$). Typical values of $\alpha$ for extended radio sources range from 0.7 to 0.8 \citep[]{Kellermann88,Condon92,PetersonBook,L&M07,Miley08,Tiwari16} and we adopt $\alpha$=0.8 as in \citet[]{Chiaberge09}, \citet[]{Gralla11}, and \citet[]{Yuan16}. We assume the radio source is at the photometric redshift of the cluster (see Table \ref{tb:vla_obs} and \S\ref{sect:spz}). 

We note that at $z$ = 1, a 1 mJy source corresponds to P$_{1.4}$ = 4.7$\times 10^{24}$ W Hz$^{-1}$. This limit is barely below the frequently quoted power boundary between the two FR classes \citep[P$_{1.4}$$\sim$ 10$^{25}$ W Hz$^{-1}$;][]{FR74,LO96}. Thus, any FR I sources we detect will be at the luminous end of the FR I distribution. 

\subsection{JVLA Follow-up Observations and Image Processing}\label{sect:vla}
\begin{deluxetable*}{cccccccccc}
    \tablecolumns{10}
	\tablecaption{Primary Radio Source Properties\label{tb:morph}}
	\tablehead{ \colhead{} & \multicolumn{5}{c}{FIRST Properties} & \colhead{} & \multicolumn{3}{c}{JVLA Derived Properties} \\
	\cline{2-6} \cline{8-10} \\
	\colhead{Cluster} & \colhead{FIRST Source} & \colhead{$R_{\mathrm{cc}}$} & \colhead{Maj. Axis} & \colhead{Int. Flux} & \colhead{$P_{1.4}$} & \colhead{} & \colhead{Morph.} & \colhead{LAS} & \colhead{Linear Size} \\ 
	\colhead{} & \colhead{} & \colhead{($^{\prime\prime}$)} & \colhead{($^{\prime\prime}$)} & \colhead{(mJy)} & \colhead{(10$^{26}$ W Hz$^{-1}$)} & \colhead{} & \colhead{} & \colhead{($^{\prime\prime}$)} & \colhead{(kpc)}}
	\startdata
	MOO J0015+0801 & J001524+080115 & 18.3 & 8.37 & 49.5 $\pm$ 0.1 & 1.85 &
	& FR II & 11.2 $\pm$ 0.88 & 86 $\pm$ 7 \\[0.2cm]
	\tableline \\
	MOO J0121$-$0145 & J012153$-$014611 & 38.4 & 9.58 & 12.1 $\pm$ 0.2 & 0.53 & 
	& UD & - & -\\[0.2cm]
	\tableline \\
	MOO J0228$-$0644 & J022802$-$064439 & 26.0 & 10.67 & 61.2 $\pm$ 0.1 & 2.06 & 
	& FR II & 16.2 $\pm$ 1.25 & 128 $\pm$ 10 \\[0.2cm]
	\tableline \\
	MOO J0250$-$0443 & J025044$-$044304 & 28.2 & 9.93 & 22.9 $\pm$ 0.1 & 1.26 & 
	& FR I & 16.5 $\pm$ 0.87 & 138 $\pm$ 7\\[0.2cm]
	\tableline \\
	\multirow{2}{*}{MOO J0300+0124} & \multirow{2}{*}{J030012+012500} & \multirow{2}{*}{34.8} & \multirow{2}{*}{7.18} & \multirow{2}{*}{10.7 $\pm$ 0.1} & \multirow{2}{*}{1.00} & 
	& UD & 7.4 $\pm$ 0.71(A) & 64 $\pm$ 6 \\
	 & & & & & &
	& UD & 4.3 $\pm$ 0.71(B) & 37 $\pm$ 6 \\[0.2cm]
	\tableline \\
	MOO J1358+2158 & J135823+215919(A) & 42.4 & 8.80 & 45.4 $\pm$ 0.2 & \multirow{2}{*}{4.79} & 
	& \multirow{2}{*}{FR II} & \multirow{2}{*}{31.6 $\pm$ 0.55} &\multirow{2}{*}{260 $\pm$ 5}\\
	 & J135822+215917(B)& 54.5 & 9.91 & 56.7 $\pm$ 0.2 & & 
	& & & \\[0.2cm]
	\tableline \\
	MOO J1412+4846 & J141255+484654 & 58.1 & 12.13 & 4.6 $\pm$ 0.2 & 0.25 & 
	& UD & 5.6 $\pm$ 1.75 & 47 $\pm$ 15 \\[0.2cm]
	\tableline \\
	MOO J1435+4759 & J143520+475953(A) & 37.0 & 7.68 & 4.0 $\pm$ 0.1 & \multirow{2}{*}{0.33} & 
	& \multirow{2}{*}{FR II} & \multirow{2}{*}{29.0 $\pm$ 1.77} & \multirow{2}{*}{240 $\pm$ 15} \\
	 & J143519+480011(B) & 53.7 & 7.80 & 2.5 $\pm$ 0.2 & & 
	& & & \\[0.2cm]
	\tableline \\
	MOO J1731+5857 & J173100+585806(A) & 34.7 & 5.52 & 1.4 $\pm$ 0.1 & \multirow{2}{*}{0.28} & 
	& \multirow{2}{*}{FR II} & \multirow{2}{*}{17.4 $\pm$ 0.72} & \multirow{2}{*}{144 $\pm$ 6} \\
	 & J173059+585804(B) & 42.9 & 8.61 & 4.1 $\pm$ 0.1 & & 
	& & & \\[0.2cm]
	\tableline \\
	MOO J2247+0507 & J224715+050806 & 18.6 & 7.48 & 332.6 $\pm$ 0.1 & 16.8 & 
	& FR II & 9.9 $\pm$ 0.70 & 82 $\pm$ 6
	\enddata
	\tablecomments{Column 3: The distance of the FIRST source from the cluster center (from FIRST catalog). Column 6: The radio luminosity calculated using Eqn.~\ref{eqn:power} and the FIRST integrated flux. Column 7: The JVLA follow-up morphology where UD denotes an undetermined morphology. If a pair of FIRST sources was correctly identified as components of a double-lobed source or if a FIRST source resolved into multiple components, we denote the components with (A) and (B). Column 8: The largest angular size (LAS) as defined in \S\ref{radio}. Column 9: The linear size is the LAS converted to kpc assuming the radio source is at the redshift of the cluster.} 
\end{deluxetable*}

\begin{deluxetable*}{cccccc}
\tablecaption{IRAC and Optical Counterpart Properties of the Radio Source \label{tb:infrared}}
\tablehead{\colhead{Cluster} & \colhead{[4.5]} & \colhead{[3.6] $-$ [4.5]} &\colhead{$i$} & \colhead{$i$ $-$ [3.6]} & \colhead{Stellar Mass} \\ & \colhead{} & \colhead{} & \colhead{} & \colhead{} & \colhead{($10^{11} M_{\odot}$)} }\\
\startdata
MOO J0015+0801 & 14.76 $\pm$ 0.03 & 0.12 $\pm$ 0.04 & 20.41 $\pm$ 0.07 & 5.53 $\pm$ 0.08 & 7.57 $\pm$ 0.21 \\
MOO J0121$-$0145 & \textgreater$~$18.69 & - & \textgreater$~$22.73 & - & \textless$~$0.2 \\
MOO J0228$-$0644 & 15.69 $\pm$ 0.03 & 0.12 $\pm$ 0.04 & 20.92 $\pm$ 0.1 & 5.11 $\pm$ 0.1 & 3.20 $\pm$ 0.09 \\
MOO J0300+0124 & 15.33 $\pm$ 0.03 & 0.41 $\pm$ 0.04 & \textgreater$~$22.73 & \textgreater$~$ 6.99 & 5.25 $\pm$ 0.14 \\
MOO J1358+2158 & 16.35 $\pm$ 0.04 & 0.32 $\pm$ 0.06 & 21.71 $\pm$ 0.16 & 5.04 $\pm$ 0.17 & 1.73 $\pm$ 0.07 \\
MOO J1435+4759 & 17.94 $\pm$ 0.18 & 0.14 $\pm$ 0.22 & \textgreater$~$22.73 & \textgreater$~$ 4.65 & 0.40 $\pm$ 0.07 \\
MOO J1731+5857 & 16.21 $\pm$ 0.04 & 0.32 $\pm$ 0.05 & \textgreater$~$22.73 & \textgreater$~$ 6.19 & 1.95 $\pm$ 0.07 \\
MOO J2247+0507 & \textgreater$~$18.69 & - & \textgreater$~$22.73 & - & \textless$~$0.2 
\enddata
\tablecomments{Magnitudes are aperture magnitudes corrected to total magnitudes. We note that MOO J0121$-$0145 is not an FR source. The uncertainty in the stellar mass does not include systematic uncertainty related to the choice of IMF, which is about 20\%. The values in this table are based on the assumptions that the counterpart is at the photometric redshift of the cluster and that AGN emission is sub-dominant to the galaxy emission at infrared wavelengths.} 
\end{deluxetable*}

\begin{figure*}
	\includegraphics[width=\linewidth, keepaspectratio]{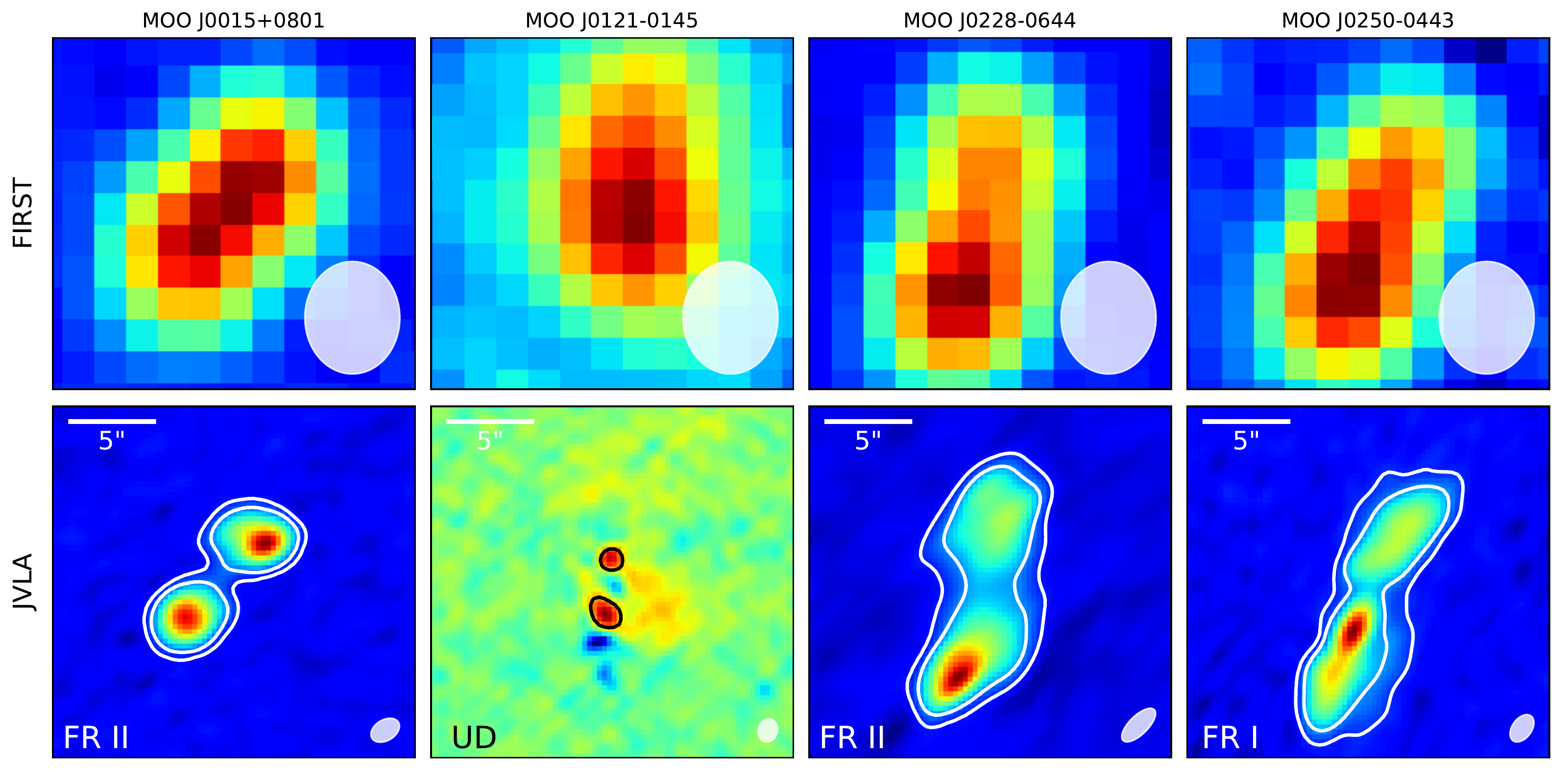}
	\includegraphics[width=\linewidth, keepaspectratio]{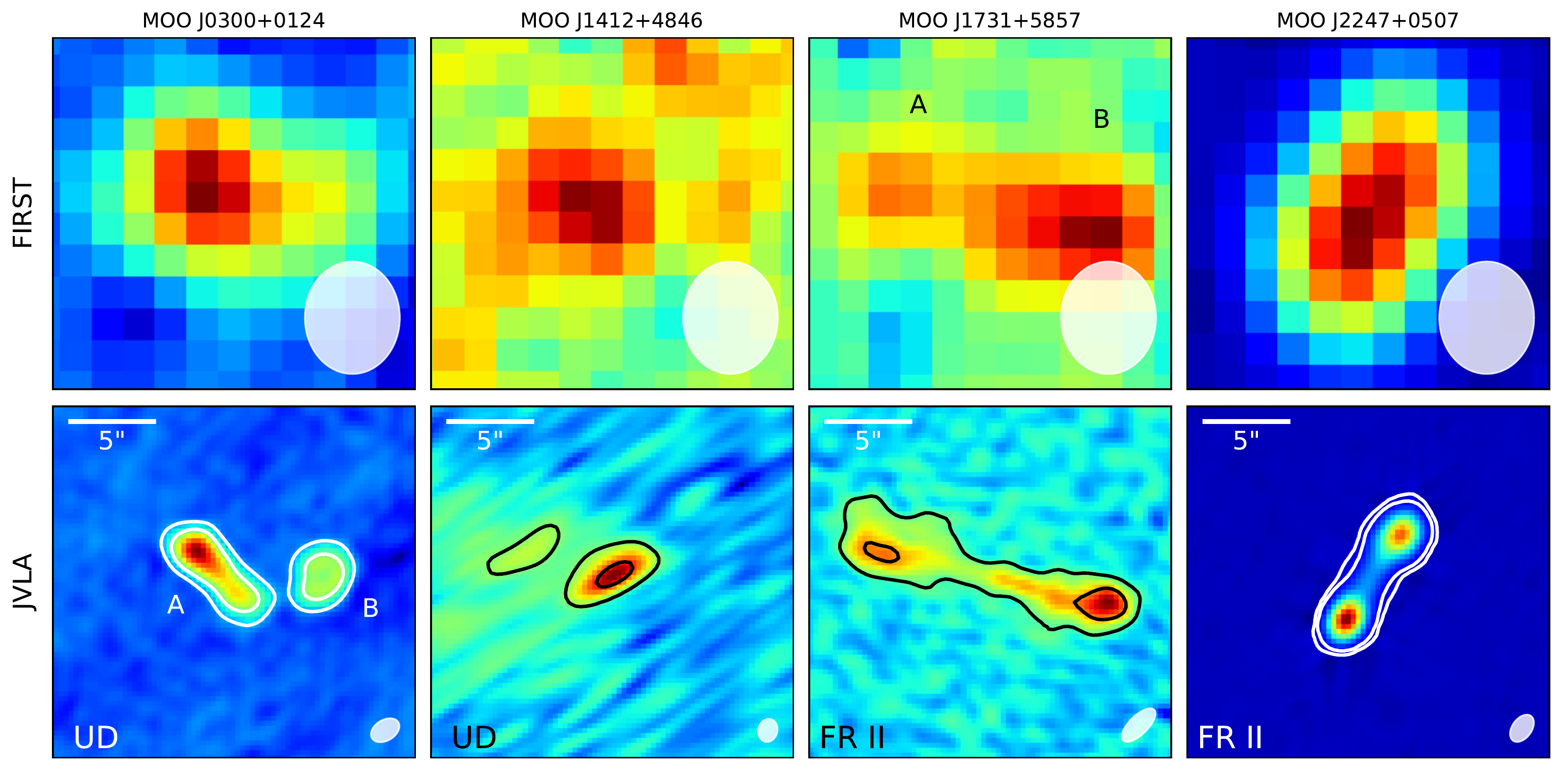}
	\caption{20$^{\prime\prime}$ $\times$ 20$^{\prime\prime}$ (160 kpc $\times$ 160 kpc) images of the targeted radio source(s) in each cluster. For each cluster, the top row is the FIRST data and the bottom row is our JVLA follow-up. North is up and East is to the left. Each image is scaled by the square root of the flux distribution. The contours levels are 4$\sigma$ and 16$\sigma$, except for MOO J0121$-$0145, MOO J0228$-$0644, and MOO J2247+2247 where the contours are 8$\sigma$ and 32$\sigma$ to showcase the morphology of the source. The synthesized beam size is shown in the lower right hand corner. The black and white contours were chosen for the guide the eye.}
	\label{fig:radio_zoom}
\end{figure*}

\begin{figure*}
    \centering
	\includegraphics[width=0.6\linewidth, keepaspectratio]{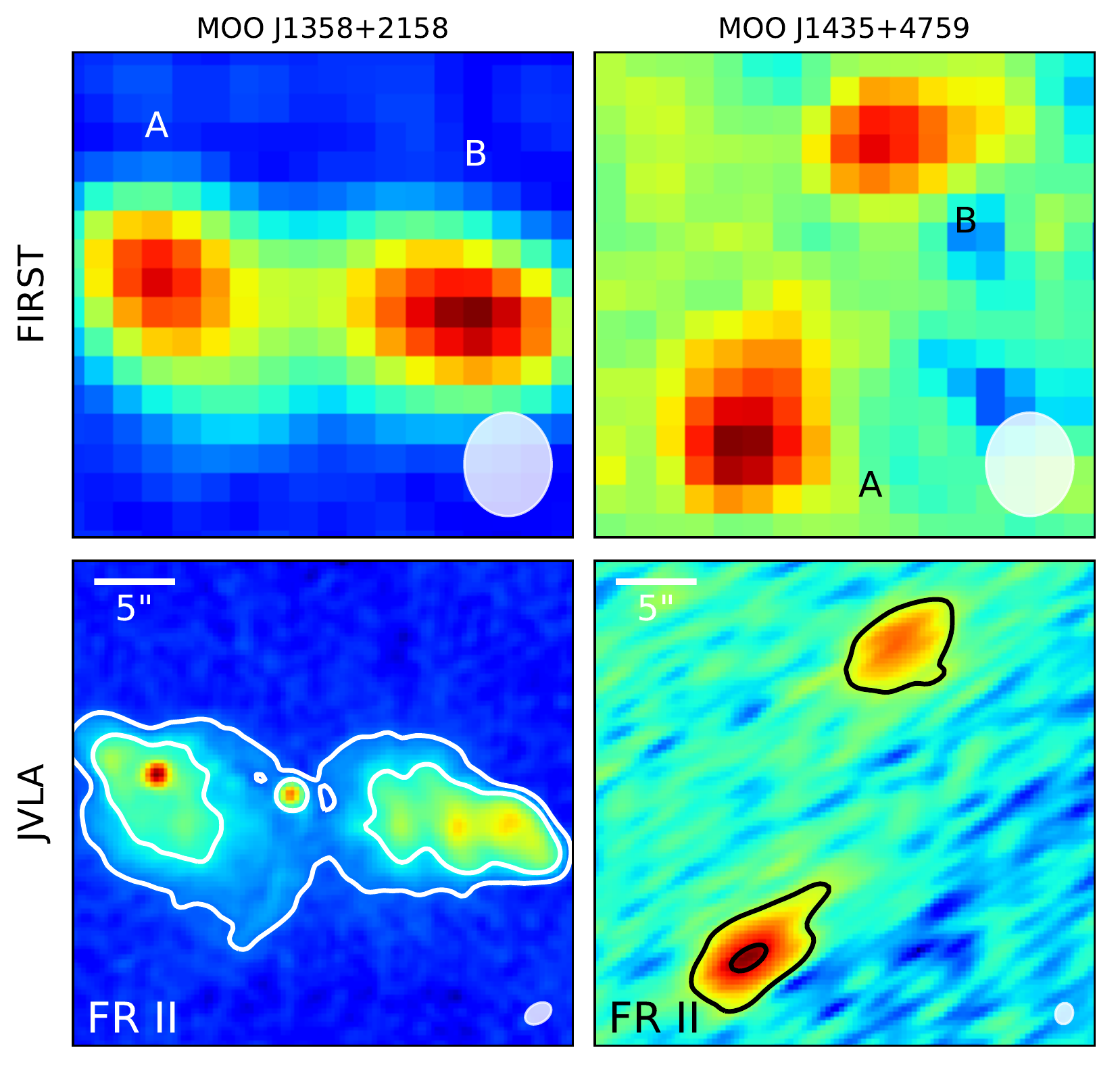}
	\caption{Same as Figure \ref{fig:radio_zoom} except the images are 30$^{\prime\prime}$ $\times$ 30$^{\prime\prime}$ (250 kpc $\times$ 250 kpc) to showcase the full extent of these sources.}
	\label{fig:radio_zoom2}
\end{figure*}

For the 2016B observing semester, we were awarded 15.5 hours of JVLA observations (PI: Gonzalez, 16B-289) to obtain high resolution imaging for a subset of the parent sample of 51 clusters. A total of 10 were observed during this semester (described in Table \ref{tb:vla_obs}) making this a pilot program for our larger ongoing program (2017B, PI: Gonzalez, 17B-197; 2018A, PI: Moravec, 18A-039). The 10 clusters observed during 2016B are representative of the larger sample. The data were taken in L-Band in the A configuration between 2016 October 26 and 2017 January 23. The observations were centered at 1.4 GHz (21cm) with a bandwidth of 600 MHz. Given this configuration and band, the resolution was $\sim$1.4$^{\prime\prime}$ using robust weighting and the primary beam was 30$^{\prime}$. The correlator was configured with 16 spectral windows, each with 64 channels. Each target was observed twice within a scheduling block for a total of $\sim$19 minutes on source.

The data were flagged, calibrated, and imaged with the Common Astronomy Software Applications (CASA) package version 4.7 \citep[]{casa}. All measurement sets were first processed through the VLA CASA Calibration Pipeline for basic flagging and calibration. We created images of each cluster by applying the TCLEAN algorithm. We cleaned the images to an average depth of $\sim$30$-$35 $\mu$Jy, which was the limit at which all obvious emission from the source was included in the clean boxes. The RMS ($\mu$Jy/beam) value for each image was determined by the following process. If possible, four rectangular regions that were near the targeted source, but did not encompass the source were drawn in the viewer. These regions were drawn in such a way as to be free of any sources. Then, these individual RMS measurements were averaged together to produce the final RMS which is reported in Table \ref{tb:vla_obs} for each image.

A pixel scale of 0.28$^{\prime\prime}$, specmode=`mfs', and a WEIGHTING = BRIGGS (robust=0.5) were used for all images. Analysis is restricted to the inner 3$^{\prime}$ of the primary beam. In the cases of MOO J0015+0801, MOO J0121-0145, MOO J0228-0644, and MOO J2247+0507, there were extremely bright sources within the field and a simple clean only partially recovered the source structure because the dirty beam was still prominent in the image. In these cases, we performed several rounds of phase-only self-calibration to increase the S/N ratio (typically by a factor of 3), reduce the prominence of improper cleaning artifacts, and recover more of the source structure.

\subsection{Spitzer Observations} \label{sect:spz}
Eight of the ten clusters were also observed during a \textit{Spitzer} Cycle 11-12 snapshot program (11080, PI: Gonzalez). Each cluster was observed for a total of $\sim$180 seconds using a set of $6 \times 30$s exposures, with a medium scale cycling dither pattern.

The catalogs were generated using a procedure similar to that of \cite{Wylezalek13} in which the Corrected Basic Calibrated Data (cBCD) was reduced and mosaicked using IRACproc (\citealt{Schuster06}), the MOPEX package (\citealt{MK05}), and a resampled pixel scale of 0\farcs6. The MOPEX outlier (e.g., cosmic ray, bad pixel) rejection was optimized for the regions of deepest coverage in the center of the maps corresponding to the position of the \madcows detection. Photometry was performed using SExtractor \citep[]{BA96} with an aperture diameter of 4$^{\prime\prime}$, corrected to total. The catalogs reach a uniform limit of 10 $\mu$Jy in [3.6] and [4.5]. 

The photometric redshifts are listed in Table \ref{tb:vla_obs} and are derived from the [3.6]$-$[4.5] and Pan-STARRS $i-$[3.6] colors of galaxies within 1$^{\prime}$ of the cluster location, which are compared with a Flexible Stellar Population Synthesis (FSPS) model \citep{c09,c10}. Details can be found in Gonzalez et al. (2018). Two clusters, MOO J0250$-$0443 and MOO J1412+4846, lack IRAC data; for these two we assume the median photometric redshift of the \madcows survey \citep[$z=1.06$,][]{G18}.

\section{Analysis \& Results} \label{results}
\subsection{Radio Morphology} \label{radio}
With the $\sim$1.4$^{\prime\prime}$ resolution of the JVLA  A-configuration at 1.4 GHz, we were able to visually determine the radio morphologies of most of the sources in our sample, which are shown in Figures \ref{fig:radio_zoom} and \ref{fig:radio_zoom2}. We visually classified the sources as either FR I, FR II, undetermined (UD) for a source with an extended morphology that is too ambiguous to categorize, or unresolved (UR) for pointlike sources. For the sources where we differed in our classifications, we discussed and agreed on a classification. In Table \ref{tb:morph}, we report the morphologies of the primary FIRST sources. If a pair of FIRST sources were correctly identified as components of a double-lobed source, we denote the components with (A) and (B).

We detect six FR II sources and one FR I source. Four clusters (MOO J0015+0801, MOO J0228$-$0644, MOO J0250$-$0443, and MOO J2247+0507) had one extended FIRST source that resolved into a double-lobed source. Three of these sources are FR IIs and one is an FR I. Three clusters (MOO J1358+2158, MOO J1435+4759, and MOO J1731+5857) had two discrete FIRST sources within 1$^{\prime}$ that were confirmed to be the constituent lobes of an FR II source. 

The radio source in MOO J1358+2158 is a canonical example of a galaxy with extended radio jets (see Fig. \ref{fig:radio_zoom2}) that have complex structures like clumps or knots. It is the only source that has a detected core. MOO J1435+4759 is the only FR radio source that has no detectable emission connecting the two lobes. Lower frequency data is needed in order to determine if the central engine has shut off or if there is emission connecting the two lobes.

The three remaining sources have more complex morphologies (see Fig. \ref{fig:radio_zoom}). In MOO J0300+0124, the FIRST source resolves into several extended sources. The eastern source could be a double-lobed source that is either viewed from an almost edge-on angle or it is a young FR I source. The western source is clearly extended towards the west. The morphological classification for either of these sources is not obvious.

In the case of MOO J0121$-$0145, the targeted FIRST source resolved into multiple compact sources, but the signal to noise is not sufficient to confidently determine the morphology. Thus we label it UD for undetermined and do not measure a size for these objects. 

In the case of MOO J1412+4846, the targeted FIRST source resolved into a compact structure with an undetermined morphology. We calculate the size of the main component. 

We note that it is possible that sources not identified as FR I or FR II could have larger scale structure that is resolved out (e.g.~MOO J0121$-$0145 \& MOO J1412+4846). Additionally, the targeted sources and secondary sources in these clusters could both be components of a common, larger scale source evidenced by the relatively close proximity of the primary and secondary sources (i.e. $\sim$ within 1$^{\prime}$) and lack of optical and infrared counterparts. But we need lower frequency or lower resolution data to conclusively determine their morphology and connection. 
\begin{figure*}
	\includegraphics[width=\linewidth, keepaspectratio]{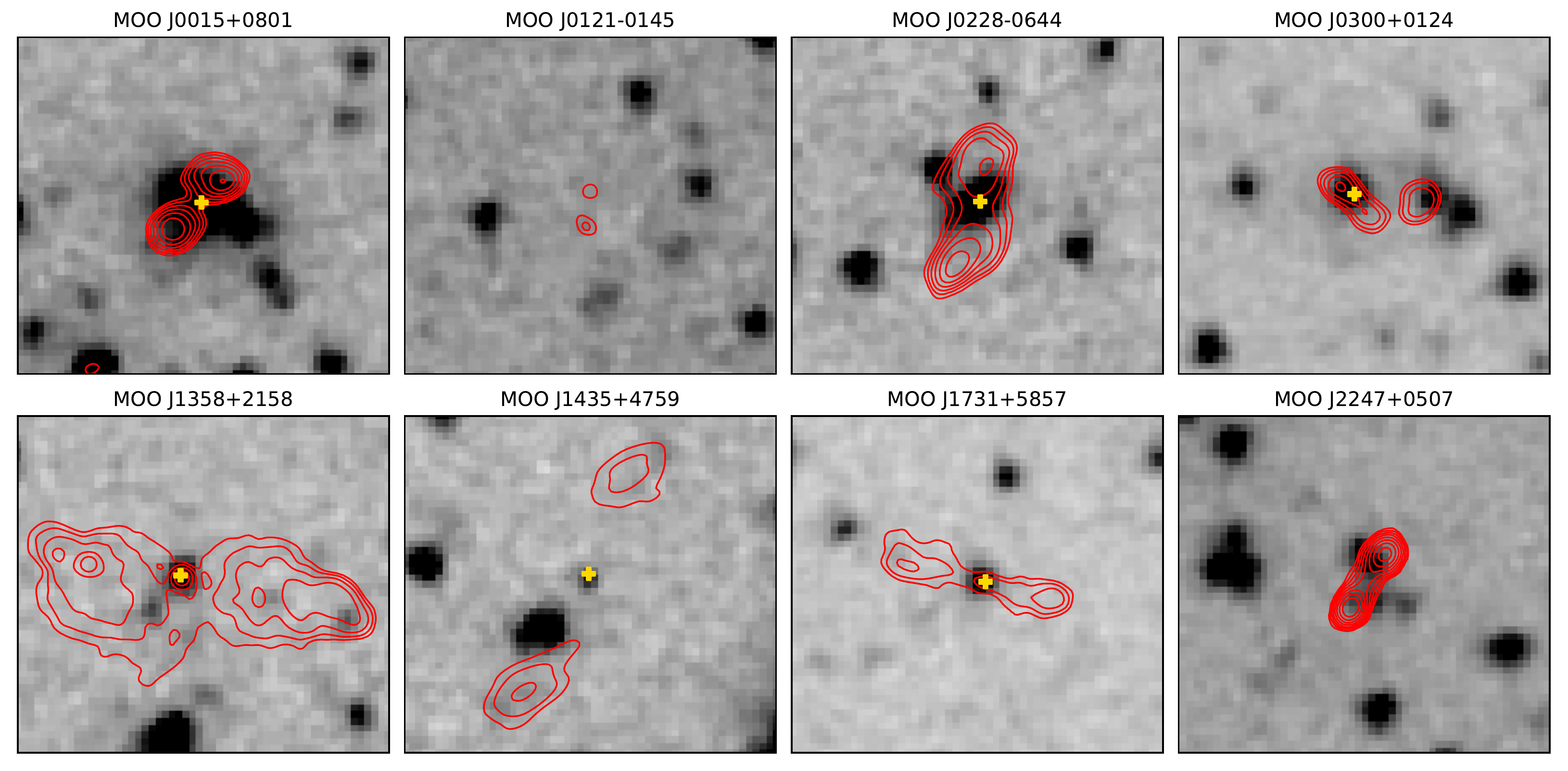}
	\caption{30$^{\prime\prime}$ $\times$ 30$^{\prime\prime}$\textit{Spitzer} 3.6$\mu$m images overlaid with JVLA contours of the radio sources. The radio contours start at 4$\sigma$ and increase by factors of 2$^n$ where n = 1,2,3, etc., except for MOO J0121$-$0145, MOO J0228$-$0644, and MOO J2247+0507 where the contours start at 8$\sigma$. Identified counterparts are denoted with a yellow plus marker. MOO J0121$-$0145 does not have an obvious infrared counterpart and MOO J2247+0507 has multiple possible counterparts.}
	\label{fig:spz_vla}
\end{figure*}

\subsection{Counterpart Analysis}
We overlay JVLA contours on the \textit{Spitzer} images for the eight clusters that have \textit{Spitzer} data and find that six of the double-lobed sources have infrared counterparts (see Figure \ref{fig:spz_vla} and Table \ref{tb:infrared}). We note that confirmation of cluster membership will require follow-up spectroscopy.

Under the assumptions that (1) these counterparts are clusters members and (2) the emission of the AGN is sub-dominant to that of the galaxy at these mid-infrared wavelengths, we can use IRAC photometry to estimate stellar masses. We convert [4.5] to a stellar mass using \verb'EzGal' \citep[]{EzGal12}. We use an FSPS model \citep[]{c09} with a simple stellar population, a formation redshift $z_f$ = 3, a supersolar metalliticity $Z=0.03$, the cluster photometric redshift, and a \cite{Chab03} initial mass function (IMF). The stellar mass changes by $\pm$20\% for $2\leq z_f \leq 5$. There is no central counterpart for MOO J0121$-$0145 or MOO J2247+0507, so we adopt a 3$\sigma$ (6 $\mu$Jy) limit to obtain upper limits on their stellar masses. We report the stellar masses in Table \ref{tb:infrared}.

We find that the stellar masses of the infrared counterparts are typically $M_*$ $\geq$ 10$^{11}$ M$_{\odot}$, while the two non-detections are constrained to have $M_*$ $\lesssim$ 2$\times$10$^{10}$ M$_{\odot}$. The majority of the counterparts appear to be fairly massive galaxies which is consistent with the expectation that extended radio lobes are predominantly hosted by massive galaxies \citep[e.g.,][]{Seymour07}. These masses are similar to the $\sim$10$^{11}$ M$_{\odot}$ masses found for BCGs at this redshift  \citep[]{Stott10,Lidman12,Bellstedt16}. The counterpart in MOO J0015+0801 is the most massive and centralized counterpart of the sample, and as such it is the most plausible candidate to be a BCG.

For those that have infrared counterparts, we calculate the observed-frame $[3.6] - [4.5]$ and $i-[3.6]$ colors (see Table \ref{tb:infrared}), using \textit{i}-band data from Pan-STARRS. In the cases of non-detections in Pan-STARRS, we assume that the counterpart has an \textit{i}-band magnitude fainter than the limiting magnitude \textit{i} = 23.1 (AB) \citep[]{PS16} and place a lower limit on the \textit{i}$-$[3.6] color. 

\begin{figure}
    \centering
	\includegraphics[width=\linewidth, keepaspectratio]{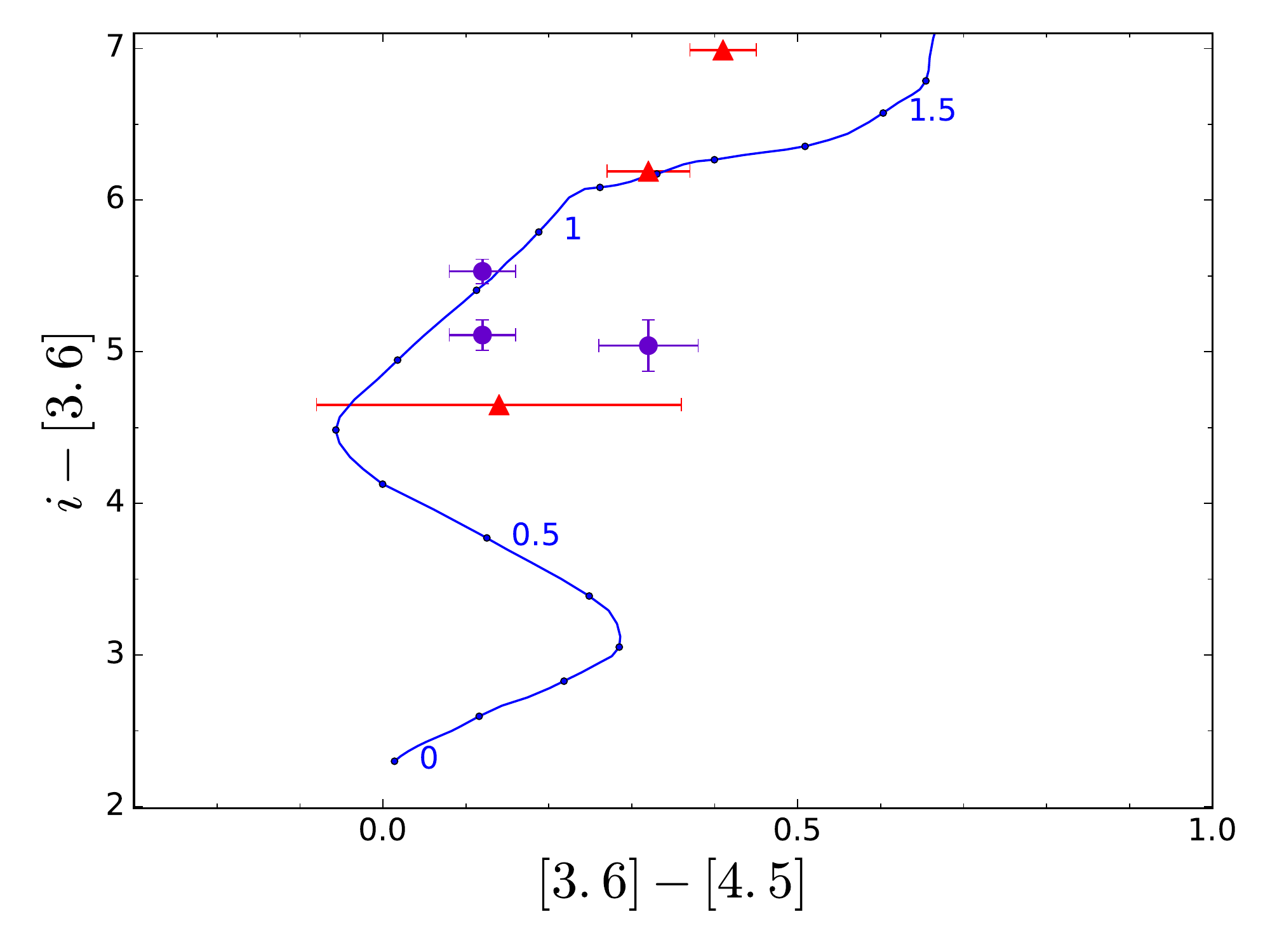}
	\caption{Color-color diagram of radio source counterparts. The blue line indicates the expected evolutionary path of a galaxy as a function of redshift which was calculated using $Z=0.03$ and the \cite{c09} models. The red triangles indicate counterparts that only have optical lower limits. The colors of most of these counterparts are consistent with the galaxy population expected in clusters at $z\gtrsim1$.}
	\label{fig:color-color}
\end{figure}

We compare the observed colors to stellar population models for passively evolving galaxies with the same parameters described above for Figure \ref{fig:color-color}. The faintness of the optical counterparts is consistent with being at the cluster redshift. We find that the colors of the counterparts are generally consistent with the galaxy population expected in clusters at this redshift for this evolutionary model. Only one galaxy, MOO J0300+0124, has a redder \textit{i}$-$[3.6] color (topmost point in Fig. \ref{fig:color-color}) than expected for a passive cluster member.

\subsubsection{Extent of Radio Sources}
The extent of the radio sources can be used as a diagnostic of the local environment. The largest angular size (LAS) is defined as the length of a straight line between the most distant points belonging to the same radio source. To determine the LAS, we measure the largest (projected) angular extent of the source contained within the lowest reliable contour. For most cases in this study the lowest reliable contour was 4$\sigma$, but for a few cases (MOO J0121$-$0145, MOO J0228$-$0644, and MOO J2247+2247) due to residual artifacts from the cleaning process, the lowest reliable contour was 8$\sigma$. These LAS measurements are given in Table \ref{tb:morph}. We convert the LAS into a largest linear (projected) size (LLS) using a scale factor that assumes the radio source is at the cluster redshift.

We investigate the relationship between the LAS and the distance from the cluster center for the FR II sources in Section \ref{discuss} where the distance of the source from the cluster center is defined as the angular distance between the center of the contours and the cluster center. We have two measurements of the cluster center which were determined by the density of the \wise and \textit{Spitzer} sources \citep[see][]{G18}. The distance of the radio source from the center of the cluster plotted in Figure \ref{fig:las_pos} is the average of the distance of the source from the {\it WISE} and {\it Spitzer} centers. The error bars were calculated by taking the difference between the two distances and dividing by two. An average error is assumed for MOO J0228$-$0644 and MOO J0250$-$0443 as MOO J0228$-$0644 does not have \textit{Spitzer} data and MOO J0250$-$0443 does not have a robust \textit{Spitzer} center. We note that these results are insensitive to precise contour measurements. If we use a constant surface brightness (derived from the noisiest image, MOO J2247+0507) at scales resolved by the JVLA instead of the standard $n*\sigma$ approach to measure the LAS, the values generally change by less than 10\%.

\section{Discussion} \label{discuss}
\begin{figure}
    \centering
	\includegraphics[width=\linewidth, keepaspectratio]{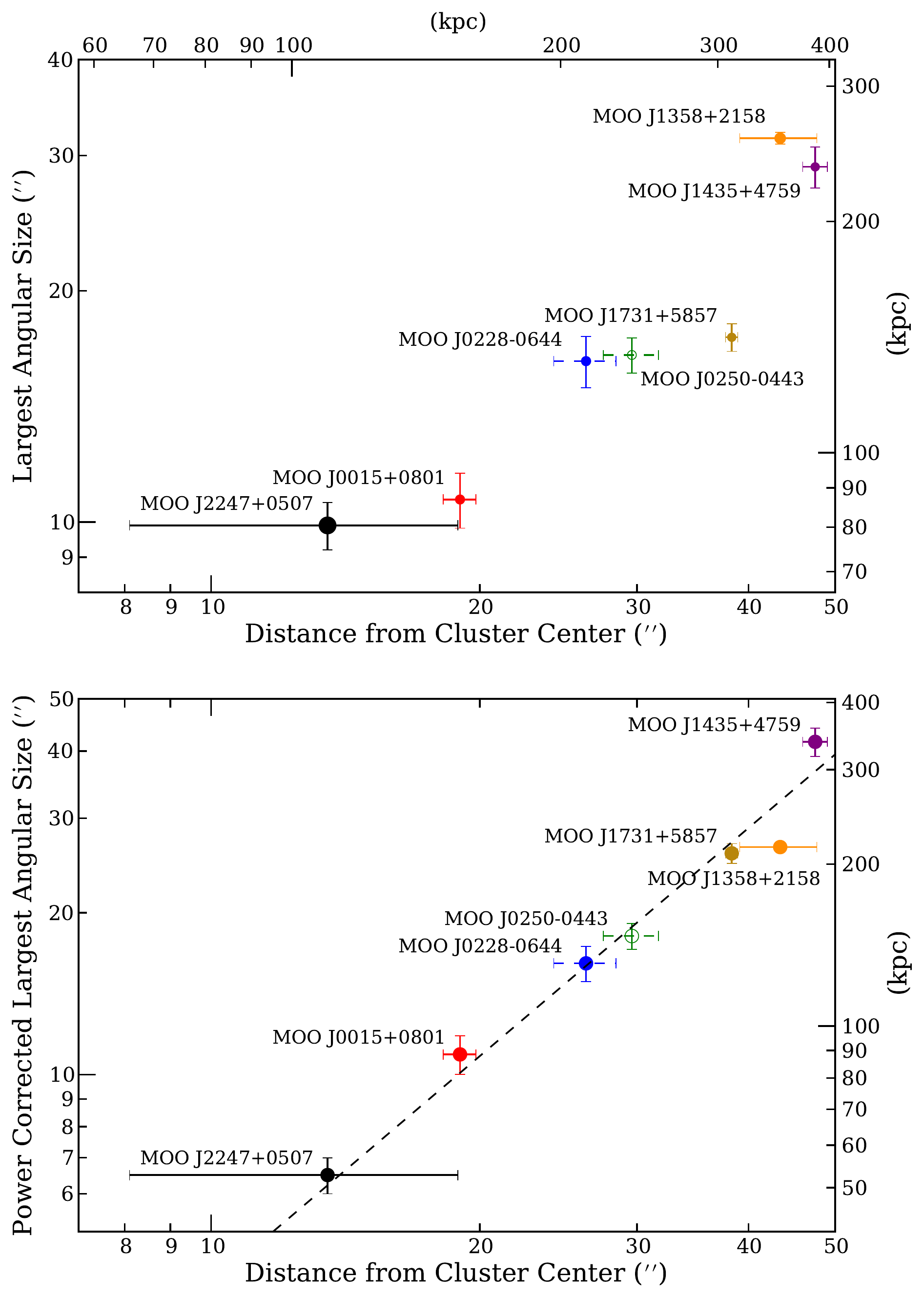}
	\caption{The relationship of largest angular size (arcseconds) versus the average distance from the cluster center. The distance is the average distance in arcseconds of the radio source to the cluster center using the \wise and \textit{Spitzer} centers (see \S\ref{radio}). The top axis is the distance from the cluster in arcseconds converted to kpc assuming $z=1$. FR II sources are marked by filled circles and the FR I source is denoted by an open circle. An average distance error is assumed for MOO J0228$-$0644 and MOO J0250$-$0443 as MOO J0228$-$0644 does not have \textit{Spitzer} data and MOO J0250$-$0443 does not have a robust \textit{Spitzer} center. This average error is depicted by a dashed error bar in both panels. \textit{Top:} The size of the marker corresponds to the radio power listed in Table \ref{tb:morph}. \textit{Bottom:} The largest angular size is normalized to a fiducial power of 2.0$\times$10$^{26}$W Hz$^{-1}$ and the relation becomes visibly tighter. A best fit (Eqn. \ref{eqn:PL_fit}) is shown with a black dashed line. All points have error bars, but in some cases the error bars are small enough that they are covered by the point itself. The tightness of the observed correlation in these plots is striking.}
	\label{fig:las_pos}
\end{figure}

We observe a strong correlation between LAS and cluster-centric radius (Figure \ref{fig:las_pos}). We begin by considering the physics driving the jet sizes. If one assumes constant power, rest-mass transport along the jet, and that the density profile of the surrounding medium is centered upon the AGN, then the functional form of the characteristic length for a self-similar FR II jet \citep{Falle91} is expected to be
\begin{equation}\label{eqn:length}
L_j = c \bigg(\frac{P}{\rho}\bigg)^{1/(5-\alpha)}t^{3/(5-\alpha)}.
\end{equation}
where $c$ is a dimensionless constant, $P$ is the power of the jet, and $t$ is the age of the source. The density of the local surrounding medium, $\rho$, is generally taken to have a profile of the form $\rho=\rho_0 r^{-\alpha}$ where $\alpha$ is the radial density index and $r$ is the distance from the source. 

While the above model is appropriate for AGN hosted by individual galaxies, if the host galaxy is in a galaxy cluster the form of the above model is only valid if the source of the jet is at the center of the gravitational potential well. If the radio source is instead not at the very center of the cluster, we can then proceed by making the simplifying assumption that there is not a strong density gradient for the local density profile (i.e. on the scales probed by the jet) centered on the radio source ($\alpha\simeq0$ and $\rho$ becomes $\rho_0$). However, $\rho_0$ will vary from source to source according to the distance from the cluster center ($R$).

In this case, Equation 2 becomes,
\begin{equation}\label{eqn:len2}
    L = c \bigg(\frac{P}{\rho_0(R)}\bigg)^{1/5}t^{3/5}.
\end{equation}
We note that while the density profile in Eqn. \ref{eqn:length} was centered on the AGN, the density profile in Eqn. \ref{eqn:len2}, represented by $\rho_0(R)$ with R being the distance from the cluster center. 

The size of the jet is thus dependent upon three physical quantities: the power of the jet, the age of the jet, and the local density of the ICM. While we have no direct measurements of the age or local ICM density, we can normalize the jet sizes to a fiducial power. In the top panel of Figure \ref{fig:las_pos}, we show the LAS versus cluster-centric radius relation. In the bottom panel of Figure \ref{fig:las_pos}, we show the same relation with the LAS now normalized to a power of $P_0$ = 2$\times$10$^{26}$W Hz$^{-1}$ using the $P_{1.4}$ from Table \ref{tb:morph} such that 
\begin{equation}
    L_{\mathrm{norm}} = L \bigg(\frac{P_0}{P_{\mathrm{1.4}}}\bigg)^{1/5}
\end{equation}
The Spearman rank coefficient between the LAS and cluster-centric radius before the power correction is $r_s = 0.96$. After the power correction, the relation between the points gives $r_s = 1.0$.

We derive a best fit assuming for the two points without \textit{Spitzer} data that the positional uncertainties are consistent with the average values for the other data points ($\pm$2.1$^{\prime\prime}$). In this case, we find
\begin{equation}\label{eqn:PL_fit}
    L_{\mathrm{norm}}= 10^{- 0.8\pm{0.6}} R^{1.4\pm{0.4}},
\end{equation} 
where $L_{\mathrm{norm}}$ is the power normalized LAS and $R$ is the cluster-centric radius. The scatter about this best fit relation is $\sigma_{log(LAS|R)}=0.047$ or 11\%. 

While physically one would expect that at larger cluster-centric radii jets should reach larger maximum sizes due to reduced pressure confinement (i.e. decreasing $\rho_0$), the tightness of this relation is striking -- particularly given the assorted factors one would expect to add scatter to the observed relation. 
Consider first simple geometric projection effects. We observe the projected cluster-centric radius and LAS rather than the true physical distance and jet size. In each case, this reliance upon projected quantities must increase the observational scatter. To illustrate the impact of this purely geometric effect, we simulate sources uniformly distributed on a spherical surface at a fixed physical radius and calculate the expected distribution of observed projected radii. The median projected radius is 87\% of the true physical radius, with a 1$\sigma$ interval spanning $0.55-0.99$. Thus, we expect an $\sim22$\% scatter ($\sigma_{geo}$) in the observed relation simply due to use of projected cluster-centric radii rather than true physical separations. 

A similar level of scatter should arise from the angle of the jet relative to the sky, with the caveat that our selection of objects with large angular size will induce some bias towards identification of FR sources oriented near the plane of the sky. While we identify FR sources oriented in the plane of the sky (since we are identifying sources by their large angular extent), nothing about our selection process induces a bias in the projected versus true physical separation from the cluster center. 

In addition to projection, there are physical factors that will impact the true length of the jet such as age and the density profile of the ICM. Given that the jet age should be uncorrelated with cluster-centric radius, a range of jet ages will increase the observed scatter -- a factor of two age difference corresponds to a factor of 1.5 change in size. The tightness of the relation requires that either all observed jets are nearly the same age, or that the time dependence is weaker than expected from the fiducial model. While the minimum size requirement does imply some bias against very young jets and jet lifetimes set an upper limit on jet length, this still leaves a relatively wide range of observable ages, and it remains implausible that all jets in this sample are the same age. 

Similarly, the size of the jet depends upon the density profile of the ICM. The presence of a tight relation with cluster-centric radius thus implies that for the clusters in our sample the ICM profiles must be similar in shape and normalization. With a larger sample it may be possible to use the distribution of jet sizes to constrain the distribution of ICM profiles. Conversely, with the addition of X-ray data the only free parameters would be jet age and system geometry.

While it is premature to draw any concrete conclusions about this tight relation due to the limited sample size, we note that this relation is much tighter ($\sigma_{LAS|D}=11\%$) than what is expected from geometric considerations alone ($\sigma_{geo}\sim22\%$). Considering the additional physical factors that should add even more scatter, this relation is surprisingly tight. \cite{Bird08} demonstrate that for a given power and age, the probability distribution function of the expected size of an FR II source of a particular age, power, and external density is strongly peaked (see their Fig. 4). This peak of the PDF provides theoretical support for the observed relation between size and cluster-centric radius being driven by the density of the surrounding medium.

\section{Summary} \label{summary}
We present the first results from our program to investigate a sample of extended radio sources associated with $z\sim1$ galaxy clusters. We have used the JVLA to obtain morphologies for the first 10 sources in our sample, combining this with \textit{Spitzer} data for eight to investigate the stellar counterparts. Our findings are the following:

\begin{itemize}
    \item{{\it Morphologies:} For this initial sample, we find that 70\% of the target radio sources exhibit a double-lobed morphology. Six of these are FR II sources and one is an FR I. The remaining three sources all resolved into multiple components, but we were unable to determine whether these were lobes or multiple radio sources.}
    \item{{\it Counterparts:} Out of the eight clusters with \textit{Spitzer} data, 75\% have infrared counterparts brighter than our detection threshold, which corresponds to $M_\star\sim2\times10^{10}$ M$_\odot$. Including non-detections, the median stellar mass is $M_\star\sim1.8\times10^{11}$ M$_\odot$, indicating that the typical host is a massive galaxy.}
    \item{{\it LAS scales with distance:} We observe a strong correlation between the largest angular extent and the distance from the cluster center.  After normalizing to a fiducial jet power, we find a scatter of only 11\% about the best fit power law relation. This level of scatter is lower than the scatter expected from purely geometric considerations, before considering the impact of variations in both jet age between radio sources and variation in ICM density profiles between clusters.}
\end{itemize}

The observations and analysis presented in this work are for the first 10 clusters observed with the JVLA as part of our larger, ongoing investigation of a sample of extended radio sources associated with 51 galaxy clusters at $z\sim1$. This is the first survey of extended radio sources in massive clusters at this redshift, and provides a unique opportunity to study the interaction between radio sources and their environment. With the larger sample we will be able to more robustly investigate the relationship between LAS scale and cluster-centric radius and the associated scatter, including full modelling of all expected sources of scatter and bias. This larger sample is also expected to yield a subset of bent tail radio sources that can be used to provide additional constraints on the surrounding ICM. Lastly, we plan to incorporate VLASS data into our analysis when the science quality data becomes available (Lacey et al. in prep).

\acknowledgements
E.M. would like to thank those at the NRAO workshops and helpdesk for invaluable help with imaging. E.M. would like to thank the anonymous referee for their helpful and insightful comments. E.M. would like to additionally thank Shuo Kong and Peter Barnes for guidance in the imaging process and Heinz Andernach for an enlightening discussion. 

This research was supported in part by the National Science Foundation (AST-1715181). This research was supported in part by NASA through the NASA Astrophysical Data Analysis Program, award NNX12AE15G. Parts of this work have also been supported through NASA grants associated with the Spitzer observations (PID 90177 and PID 11080). Basic research in radio astronomy at the U.S. Naval Research Laboratory is supported by 6.1 Basic Research. 

This publication made use of Common Astronomy Software Applications (CASA) package \citep[]{casa}.~This publication made use of Astropy, a community-developed core Python package for Astronomy \citep[]{astropy} and APLpy, an open-source plotting package for Python \citep[]{aplpy}.

This publication makes use of data products from NRAO's Karl G. Jansky Very Large Array (VLA). The National Radio Astronomy Observatory is a facility of the National Science Foundation operated under cooperative agreement by Associated Universities, Inc. Additionally, this publication makes use of data products from the Wide-field Infrared Survey Explorer, a joint project of the University of California, Los Angeles, and the Jet Propulsion Laboratory/California Institute of Technology, funded by NASA. This work is also based in part on observations made with the Spitzer Space Telescope, which is operated by the Jet Propulsion Laboratory, California Institute of Technology under a contract with NASA. 

%%%%%%%%%%%%%%%%% APPENDICES %%%%%%%%%%%%%%%%%%%%%
\appendix
\section{Secondary Sources}\label{second_sources}
 
\begin{figure*}
    \centering
	\includegraphics[width=0.8\linewidth, keepaspectratio]{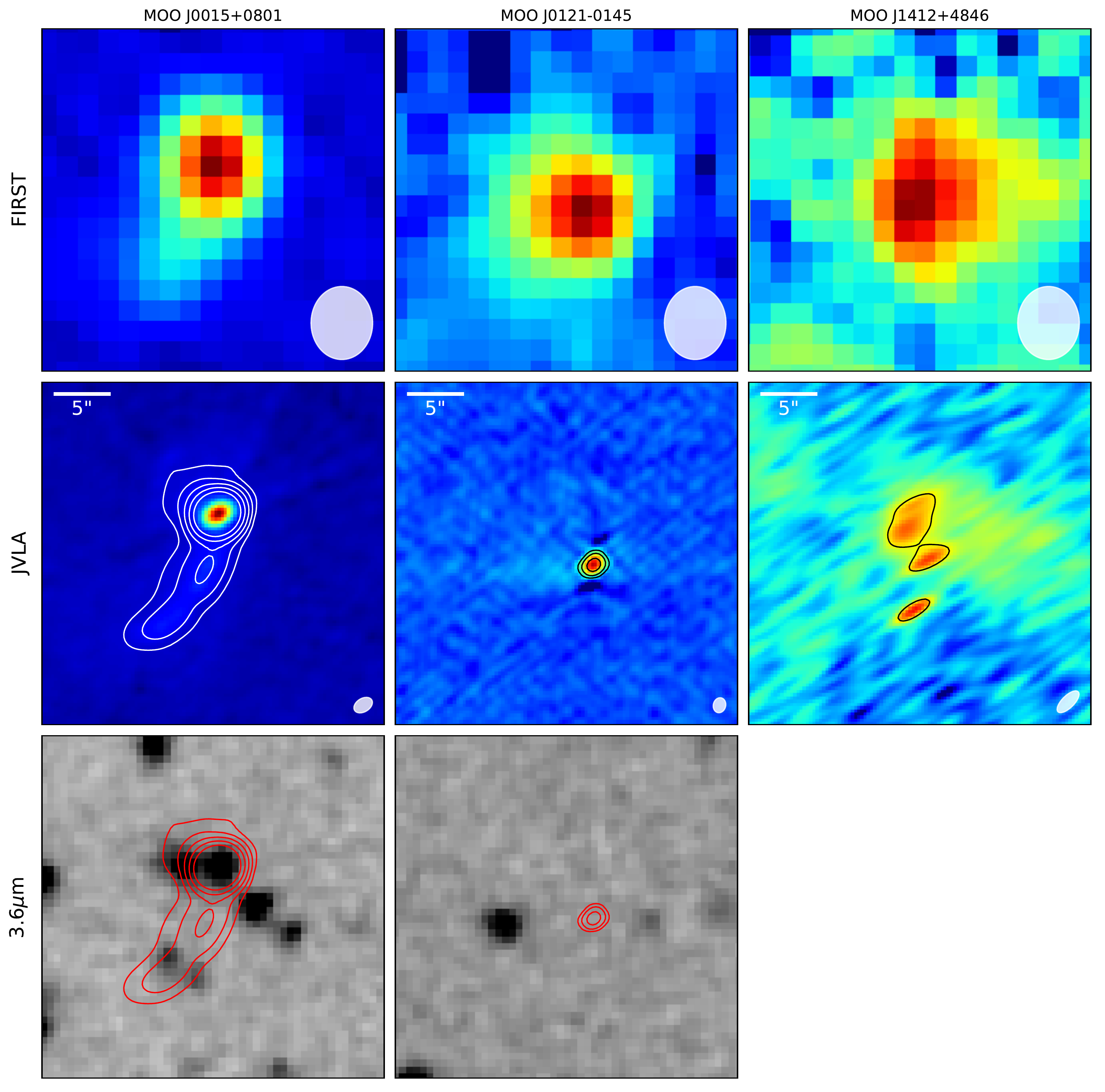}
	\caption{20$^{\prime\prime}$ $\times$ 20$^{\prime\prime}$ ($\sim$ 160 kpc) images of secondary FIRST sources. For each cluster, the top image is the FIRST radio source, the middle image is the JVLA source, and the bottom image is the \textit{Spitzer} 4.5$\mu$m image. North is up and East to the left. The contours levels increase by powers of two starting at 8$\sigma$. For MOO J0015+0801 the contours are smoothed. The synthesized beam size is shown in the lower right hand corner.}
	\label{fig:secondary}
\end{figure*}

MOO J0015+0801 has a second FIRST source in addition to the double-lobed source (see Fig. \ref{fig:secondary} and Table \ref{tb:morph_add}). The morphology of this source is interesting as the southern portion resembles an FR I bent-tail source, whereas the northern portion of this source is less elongated than simply extended, with only a small indication of bending and does not fit into an obvious morphological classification. If assumed to be a bent-tail source, it has an opening angle of $\sim$115$^{\circ}$ and is thus classified as a Wide Angle Tail (WAT) source. The opening angle was calculated using the distances from the extent of the 4$\sigma$ contour along the jet-axis to the core and applying spherical geometry. An infrared counterpart is detected in the \textit{Spitzer} image.

In the case of MOO J0121$-$0145 (Fig. \ref{fig:secondary}), the targeted FIRST source is unresolved in the JVLA data. There are characteristic artifacts from wide-field wide-band imaging in the JVLA image, but this was not the primary source of interest. An infrared counterpart is not detected in the \textit{Spitzer} image. 

In the case of MOO J1412+4846 (Fig. \ref{fig:secondary}), the signal to noise is not sufficient to confidently determine the morphology of the targeted FIRST source. Thus we label it UD in Table \ref{tb:morph_add}.

\begin{deluxetable*}{cccccc}
	\tablecaption{Secondary Radio Source Properties\label{tb:morph_add}}
	\tablehead{\colhead{Cluster} & \colhead{FIRST Source} & \colhead{R$_{cc}$} & \colhead{Maj. Axis} & \colhead{Int. Flux} & \colhead{Morph.} \\ & & \colhead{($^{\prime\prime}$)} & \colhead{($^{\prime\prime}$)} & \colhead{mJy} & }
	\startdata
	MOO J0015+0801 & J001525+080203 & 42.7 & 4.16 & 84.7 $\pm$ 0.1 & UD\\
	\tableline \\
	MOO J0121$-$0145 & J012149$-$014557 & 34.2 & 5.96 & 11.7 $\pm$ 0.1 & UR \\
	\tableline \\
	MOO J1412+4846 & J141259+484703 & 65.8 & 8.82 & 5.9 $\pm$ 0.2 & UD \\
	\enddata
	\tablecomments{Column 3: The distance of the FIRST source from the cluster center (from FIRST catalog). Column 6: The radio luminosity calculated using Eqn. \ref{eqn:power} and the FIRST integrated flux. Column 7: The JVLA follow-up morphology where UD denotes an undetermined morphology.} 
\end{deluxetable*}

%%%%%%%%%%%%%%%%%%%%%%%%%%%%%%%%%%%%%%%%%%%%%%%%%%
%%%%%%%%%%%%%%%%%%%%%%%%%%%%%%%%%%%%%
\bibliography{bibliography_RadCows}{}
\bibliographystyle{apj}

\end{document}